\documentclass{eas}
\usepackage{graphicx}
\def \cm{~\rm{cm}}
\def \s{~\rm{s}}

\def \yr{~\rm{yr}}
\def \pc{~\rm{pc}}
\def \kpc{~\rm{kpc}}

\begin{document}

\title{THE ROLES OF JETS:  CF, CCSN, PN, CEE, GEE, ILOT}
\runningtitle{The roles of jets}
\author{Noam Soker}\address{Department of Physics, Technion -- Israel Institute of Technology, Haifa 32000 Israel; soker@physics.technion.ac.il}
\begin{abstract}
I review the roles of jet-inflated bubbles in determining the
evolution of different astrophysical objects. I discuss
astrophysical systems where jets are known to inflate bubbles
(cooling flow [CF] clusters; young galaxies; intermediate
luminosity optical transients [ILOTs]; bipolar planetary nebulae
[PNe]), and systems that are speculated to have jet-inflated
bubbles (core collapse supernovae [CCSNe]; common envelope
evolution [CEE]; grazing envelope evolution [GEE]). The jets in
many of these cases act through a negative jet feedback mechanism
(JFM). I discuss the outcomes when the JFM fizzle, or does not
work at all. According to this perspective, some very interesting
and energetic events owe their existence to the failure of the
JFM, including stellar black holes, gamma ray bursts, and type Ia
supernovae.
\end{abstract}
\maketitle

\section{Introduction}
 \label{sec:intro}

In many astrophysical objects jets play significant roles in the
evolution. In the present study it is assumed that when ever an
accretion disk (or a belt) with a sufficiently high accretion rate
is formed, two opposite jets are launched.

In section \ref{sec:ibubbles} I list several groups of
astrophysical objects where bubbles and lobes are inflated by
jets. This is an updated version of the list presented by Soker
{\em et al.} (\cite{Sokeretal2013}). I do not list objects where
jets typically do not inflate bubbles. These objects include young
stellar objects and some elliptical planetary nebulae (PNe), both
of which are old members of astrophysical objects with jets, and
the recently proposed group of Type Ia supernova remnants
(Tsebrenko \& Soker \cite{TsebrenkoSoker2013}).

In some of the above listed objects with jet-inflated bubbles, the
operation of the jets is regulated by a negative jet feedback
mechanism (JFM). In section \ref{sec:fizzle} I list those objects
and some of their properties that are related to the JFM. I then
discuss and speculate on the outcomes when the JFM fizzle at early
times, or does not work at all. In this 4-pages-limited paper I
only present the fizzle-outcomes; a more extended discussion will
be presented elsewhere.

\section{Inflating bubbles}
 \label{sec:ibubbles}

Table 1 that is based on the one presented by Soker {\em et al.}
(\cite{Sokeretal2013}), presents the groups of objects that were
claimed to have jet-inflated bubbles. Detail comparison of
morphologies is conducted by Soker {\em et al.}
(\cite{Sokeretal2013}). The table is updated with the groups of
grazing envelope evolution (GEE; Soker \cite{Soker2015}), and that
of intermediate luminosity optical transients (ILOTs) that were
added by Kashi \& Soker (\cite{KashiSoker2015}) to the groups of
objects where the JFM operates. In the GEE the jets launched by
the companion are very efficient in removing the giant envelope
gas, such that the companion never gets deep into the envelope. It
can spirals-in, but no envelope will be left outside its orbit. It
is possible that a GEE will turn to a common envelope evolution
(CEE) if the jets do not manage to eject the envelope, and the
companion spirals-in deep into the giant envelope. The suggestion
that the JFM might operate in some ILOTs is based on the
high-accretion-powered ILOT (HAPI) model (Kashi \& Soker
\cite{KashiSoker2015}). In the HAPI model it is assumed that ILOT
events are powered by accretion of mass at a very high rate onto a
main sequence (MS), or slightly evolved off the MS, star.
\begin{table*}
\tiny
\caption{Properties of jet-inflated bubbles}
    \begin{tabular}{l l l l l l l l}
     \hline
                     & Clusters           & Young     & CCSNe            & Bipolar        & CEE               &GEE                   & ILOTs      \\
                      & CFs               & galaxies  &                  & PNe            &                   &                      &            \\
         \hline
         \hline \vspace{0.00cm} \\
        Energy (erg) &     $10^{60}$      &   $10^{59}$     & $10^{51}$        &$10^{44}$       & $10^{44-48}$ & $10^{44-48}$    &    $10^{44-50}$     \vspace{0.1cm}\\
        Mass $(M_\odot)$& $10^{12}$       &  $10^{11}$      & $10$             &$1$             & $1$               & $1$                  & 1-100    \vspace{0.1cm} \\
        Size         & $100 \kpc$         & $10 \kpc$       & $10^9 \cm$       &$0.1 \pc$       & $10^{1-2} R_\odot$  & $10^{1-2} R_\odot$ & $10^{1-2} R_\odot$        \vspace{0.1cm}\\
        Time         & $10^{7-8} \yr$    & $10^{7-8} \yr$ & $1-3 \s$         &   $10^{1-2} \yr$   & $1 \yr$           & $100 \yr$            & days-yrs  \vspace{0.1cm}\\
        $T_{\rm bubble}(K)$ & $10^{9-10} $    & $10^{9-10}$  &  $10^{10} $      &   $10^6 $      & $10^{7-9}$       & $10^{7-9}$          &  $10^{6-8}$   \vspace{0.1cm}\\
        $T_{\rm res}(K)$& $10^{7-8} $       &  $10^{6-7} $   & few$\times 10^9$ & $10^4 $        & $10^{5-6}$   & $10^{5}$             &  $10^{4-6}$    \vspace{0.1cm}\\
        Accretor        & SMBH                  & SMBH             & NS/BH         & MS/WD       & MS/WD/NS          & MS/NS             &  \vspace{0.1cm} \\
            $M_a$ $(M_\odot)$ & $10^{8-10}$& $ 10^{6-9} $  & $1-5 $           & $1$             &$1$                &$1$                & 1-50         \vspace{0.1cm}\\
     \hline
        Jets' main    & Heating            & Expelling      & Exploding       &Shaping          & Ejecting             & Ejecting                     & Shaping;       \\
        effects       & ICM                & gas        & the star      &lobes            & envelope          & outer                     & radiation        \\
                      &                    &            &               &                 &                   & envelope                  &               \vspace{0.1cm} \\
     \hline
        Observation     & X-ray        & Massive      &               & Bipolar       &                       &                & Bipolar      \\
                        & bubbles      & outflows     &              & PNs            &                       &                & LBVs      \\ \hline
    \end{tabular}
\label{tab:Table1}
\begin{flushleft}
\small Properties of systems where jet-inflated bubbles are
observed or assumed to exist. The different listed values are
typical and to an order or magnitude (or two even) accuracy only.
The quantities given are as follows. Energy in one jets-launching
episode; system mass; size of the relevant ambient gas; duration
of the jets activity episode; temperature of the gas inside the
bubble; temperature of the ambient gas (the subscript `res' stands
for reservoir of gas for accretion); the accreting compact object
that launches the jets; its mass; the main effect of the jets;
resolved observed outcomes of the jets activity.
 \newline
 Acronym: \textbf{BH}: black hole; \textbf{CCSNe}: core
collapse supernovae; \textbf{CEE}: common envelope evolution;
\textbf{CFs}: cooling flows; \textbf{GEE}: grazing envelope
evolution; \textbf{ICM}: intra-cluster medium; \textbf{ILOT}:
intermediate luminosity optical transient; \textbf{LBV}: luminous
blue variable; \textbf{MS}: main sequence star; \textbf{NS}:
neutron star; \textbf{PNe}: planetary nebulae; \textbf{SMBH}:
super massive BH.
    \end{flushleft}
\end{table*}
In this short article I only mention the following in regards to
Table 1.
\newline
(1) \emph{Observations.} Bubbles are clearly observed in cooling
flows (CFs) in galaxies and in clusters of galaxies, in bipolar
planetary nebulae (PNe), and in some ILOTs, such as the Great
Eruption of $\eta$ Car that formed the Homunculus in the $19^{\rm
th}$ century. As well, jets are known to be active in galaxy
formation.
\newline
(2) \emph{{Expelling gas.}} Jets are thought to expel gas in the
process of galaxy formation. In some other cases jet-inflated
bubbles are hypothesized to eject mass. In light of the failure of
neutrino-based processes to explode massive stars, the
jittering-jets model to explode massive stars as core collapse
supernovae (CCSNe) was proposed. Jets are hypothesized to
facilitate mass ejection in many cases of common envelope
evolution (CEE). Jet are very efficient in expelling the giant
envelope gas outside the secondary orbit in the GEE.  While jets
are crucial in CCSNe and in the GEE according to this picture,
they are not mandatory in the CEE.
\newline
(3) \emph{The JFM.} The JFM is a crucial in some systems, while in
some systems, like the shaping of bipolar PNe by jets, the JFM
does not operate. In the CEE and in ILOTs the JFM might act, but
it does not have a crucial role. In galaxy formation (young
galaxies), in CFs, and in CCSNe the JFM plays a crucial role.
Table 2 lists the role of the JFM in the different systems. In
shaping bipolar PNe it has no role, and hence bipolar PNe are not
listed in Table 2. One should notice the difference between the
role of the jets, as listed in Table 1, and the role of the JFM in
regulating the activity cycle, the duration, and the power of the
jets.
\begin{table*}
\tiny
\caption{JFM properties and its fizzle outcomes}
    \begin{tabular}{l l l l l l l l}
     \hline
                                      & Clusters       & Young          & CCSNe         & CEE$^{[1][2]}$& GEE$^{[1]}$   &     ILOTs$^{[2]}$      \\
                                      & CFs            & galaxies       &               &               &               &            \\
         \hline
         \hline                                                                                                                              \vspace{0.00cm} \\
 Accretor                             & SMBH           & SMBH           & NS/BH            & MS/WD/NS      & MS/NS            &     MS             \vspace{0.1cm} \\

 $R_a$(cm)                            & $10^{13-16}$   & $10^{11-14}$   & $10^{6}$      & $10^{11}$     & $10^{11}$     &    $10^{11-12}$    \vspace{0.1cm} \\

$\Phi_a ({\rm km}/{\rm s})^2$         & $ c^2 $        &  $ c^2 $       & $(10^5)^2$    &  $(500)^2$    & $(500)^2$     &    $(1000)^2$      \vspace{0.1cm} \\

 $R_{\rm res}$(cm)                    & $10^{21-24}$   & $10^{20-23}$   & $10^9$        & $10^{13}$     & $10^{13}$     &    $10^{11-13}$    \vspace{0.1cm} \\

 $\Phi_{\rm res}({\rm km}/{\rm s})^2$ & $(1000)^2$     & $(300)^2$      & $(10\,000)^2$ & $(30)^2$      & $(30)^2$      &    $(30-10^3)^2$   \vspace{0.1cm} \\

 $\Phi_a/\Phi_{\rm res}$              & $\approx 10^5$ & $\approx 10^6$ & $\approx 100$ & $\approx 100$ & $\approx 100$ &    $\approx 3-100$ \vspace{0.1cm} \\
     \hline
 Role of                              & Maintain       & SMBH-bulge     & Explosion       & Not    &  Ensures           & Might limit                                    \\
 the JFM                              & ICM            & mass           & energy $\approx$& much   &  outer             & accretion rate,                 \\
                                      & temperature    & correlation    & binding         &        &  envelope          & but much above                   \\
                                       &               &                & energy          &        &  removal           & Eddington limit                 \vspace{0.1cm}  \\
     \hline
 Fizzle                               & Cooling        & Rapid          & BH              & Core-         & Forming     &                           \\
 outcome                              & catastrophe    & SMBH           & formation;      & secondary     & a common    &                            \\
                                      &                & growth         & GRB             & merger        & envelope    &                    \vspace{0.1cm}  \\
\hline
    \end{tabular}
\label{tab:Table2}
\begin{flushleft}
\small The role of the jet feedback mechanism (JFM). $R_a$ and
$\Phi_a$ stand for the typical radius of the accretor, and the
gravitational potential on its surface. $R_{\rm res}$ and
$\Phi_{\rm res}$ stand for the typical radius of the reservoir of
gas for accretion and the energy required to expel it from the
system.
\newline
\small \footnotemark[1]{We refer here only to a giant primary star
(AGB, RGB, etc.) }
  \newline
 \footnotemark[2]{While in the other objects jets are crucial ingredient in the evolution, in the CEE and in ILOTs in some cases jets do not occur, or play a small role.
 As well, not in all cases the JFM operates even if jets do exist.}
    \end{flushleft}
\end{table*}

\section{When the jet feedback mechanism fizzle}
 \label{sec:fizzle}

I discuss the outcomes when the JFM fizzle. One way by which the
JFM can fail is if the jets are well collimated and preserve a
non-variable axis, hence the jets interact with a small fraction
of the reservoir gas along the polar directions; mass is
continuously accreted from the equatorial plane. This is the case
when the angular momentum of the accreted gas has a well-defined
direction. The jets cannot stop the accretion even when their
power increases.

In some systems the JFM is still a speculative process not in the
consensus, and so the fizzle-outcomes in these systems will be
speculative as well. These systems include the CEE, GEE and the
jittering-jets model in CCSNe.
\newline
 \textbullet \emph{Clusters CFs.}
When the AGN jets do not heat the ICM efficiently, a cooling
catastrophe might occur. The ICM cools on its radiative cooling
time, and leads to high rate star formation. Mass feeds the SMBH
as well, and eventually energetic jets will heat the ICM to
reestablish the JFM.
\newline
 \textbullet \emph{Young Galaxies.} The JFM can fail in preventing the growth
of the SMBH if the accreted gas posses a well defined angular
momentum axis over a long time. This might be the case in
elliptical galaxies or bulges that posses fast and regular
rotation. I suggest this explanation to the over-massive SMBHs in
the galaxies MRK1216 and NGC1277 that have over-massive SMBHs and
have fast and regular rotation (Y{\i}ld{\i}r{\i}m {\em et al.}
\cite{Yildirim2015}).
\newline
 \textbullet \emph{CCSNe.} The JFM mechanism can be inefficient in CCSNe when
the pre-collapse core is rapidly rotating. In this case a well
defined accretion disk is formed around the newly formed NS, and
the jets have a well define propagation axis. The jets do not
eject much of the core gas near the equatorial plane. A BH might
formed that launches relativistic jets. This is the scenario for
gamma ray bursts (GRB) in the jittering-jets model. Eventually the
jets do remove large portion of the stellar mass, and a CCSN does
take place in parallel to the GRB.
\newline
 \textbullet \emph{CEE.} Jets might not be necessary to eject the envelope in
all CEE cases. In many cases where jets are not efficient,
however, I speculate that merger of the secondary star with the
core takes place. This might especially be the case with WD
companions spiraling inside the envelope of a relatively massive
giant, more than about $3 M_\odot$. The merger product might be a
Type Ia supernova progenitor according to the core-degenerate
scenario.
\newline
 \textbullet \emph{GEE.} The GEE is based on efficient envelope gas removal by
jets. This prevents the formation of a CE. When the jets fail in
that, a CEE will commence.


\end{document}